# Optical excitation of surface plasmons and terahertz emission from metals

I.V. Oladyshkin, D.A. Fadeev and V.A. Mironov

oladyshkin@gmail.com

Institute of Applied Physics of the Russian Academy of Sciences, 603950, Nizhny Novgorod, Russia

We propose a microscopic theory of terahertz (THz) radiation generation on metal gratings under the action of femtosecond laser pulses. In contrast to previous models, only low-frequency currents inside the metal are considered without involving electron emission. The presented model is based on plasmon-enhanced thermal effects and explains the resonant character of optical-to-THz conversion giving an adequate estimation for the full signal energy. Numerical modeling reproduces specific experimental features like delayed character of THz response and low conversion efficiency when the grating depth is too large.

Plasmonics is an intensively developing field covering optics and condensed matter physics. The basic concept connecting these two areas is "plasmon polariton" – collective oscillation of free electrons coupled with the oscillation of electromagnetic field.

One of the most significant achievements of plasmonics is the possibility to localize light in subwavelength area, for instance, near the metal surface or around the nanoparticle. If coupling between some plasmonic mode and the optical wave in free space is efficient enough, the electrical field grows by orders of magnitude. This opens up new possibilities to investigate nonlinear optical properties of materials using relatively low incident radiation intensities. To date, many aspects of plasmonics have been studied experimentally and theoretically: plasmon-enhanced generation of optical harmonics [1], thermo-plasmonics [2], acceleration of electrons near the structured metal surface [3], plasmonic-induced photochemistry [4] and many others.

In this Letter we will focus on the optical excitation of surface plasmons and terahertz (THz) radiation generation in metals under the action of femtosecond laser pulses. Despite the detailed experimental study performed over the past ten years [5-9], the microscopic mechanism of conversion of surface plasmons (SP) at optical frequency to the electromagnetic pulse at THz frequency is still unclear. The main interest to this nonlinear effect is mostly fundamental, since the conversion process probably involves not well-studied features of SP at subpicosecond timescale.

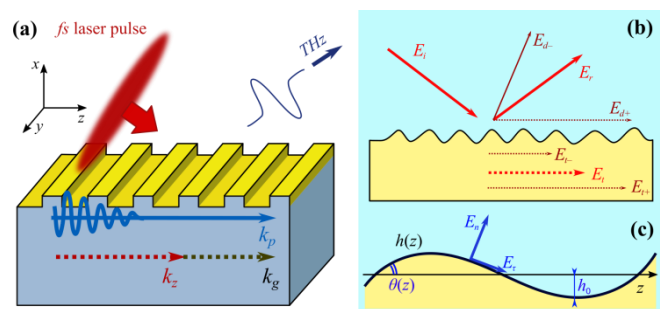

Fig 1. (a) Typical experimental scheme of SP-enhanced THz generation: irradiation of a gold foil deposited on a substrate by femtosecond laser pulse. Resonant conditions for the forward surface plasmon excitation is $k_z + k_g = k_p$ (the algebraic sum of grating wavenumber and "z" component of optical wavevector equals to the surface plasmon wavenumber). (b) Electromagnetic waves diffraction on the grating. Arrows show the wavevectors of the incident, reflected and diffracted waves, bold (thin) style of lines denotes waves of zero (first) order of perturbation theory, dotted style denotes waves which are localized near the surface in "x" direction. (c) Boundary conditions at the interface.

Plasmon-enhanced THz generation from metal was first observed in 2007 by Welsh et al. in the experiments with gold gratings irradiated by 100-fs laser pulses at different angles of incidence [5]. It was found that THz response of a grating resonantly increases when the condition of SP excitation is fulfilled (algebraic sum of grating wavenumber and tangential component of optical wave vector equals to plasmon wavenumber – see Fig. 1 (a) and Eq. (11) below). Resonant nature of optical-to-THz conversion was also confirmed in later experiments [6-9].

In their papers [5, 6] Welsh et al. suggested that the THz response was produced by the electrons emitted in free space and accelerated by the ponderomotive potential of SP field. However, later experiments [7, 8] showed that a structured metal foil emits roughly the same THz signal being irradiated from both vacuum side and substrate side. Since free electron acceleration into the substrate is impossible, the discussed model cannot interpret the fact of bi-directional generation. In recent measurements at low laser fluences [9] the quadratic power law of the conversion process was confirmed which also contradicts the mechanism based on the electron emission.

Here we propose a free-electron model of THz response based on low-frequency current excitation by fast temperature gradient inside the electronic subsystem of metal. This mechanism was first considered for flat metal surfaces in [10] and allowed to interpret most of experimental features like the THz signal energy, significant delay of the THz response, dependence of the conversion efficiency on the laser pulse polarization and others (for details see [11-14]). In this Letter we generalize the thermal model of optical-to-THz conversion taking into account the possibility of SP excitation. In the first part of the Letter we will develop an analytical theory of SP excitation and absorption for the case of shallow grating; the second part is devoted to numerical modeling and comparison with the experimental results.

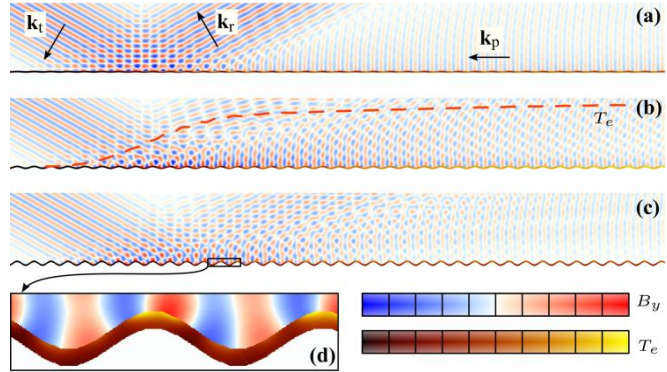

Fig 2. Magnetic field at optical frequency $B_y$ above the metal foil (red and blue colors) and distribution of the electronic temperature inside it (yellow to black) obtained in the numerical modeling. (a) The case of shallow grating. Wavevectors of the SP, incident and reflected pulses are shown. (b) Optimal grating depth (see Fig. 4 below); electronic temperature increase is shown by the dotted line. The field behind the laser pulse is formed by the interference of SP and diffraction wave $\mathbf{E}_{t-}$ (c) Large grating depth; inset (d) shows the near-field and temperature distribution.

To study analytically the contribution of SP to the total heating of electrons we need to derive an explicit formula connecting the amplitude of SP and the amplitude of the incident optical pulse $\mathbf{E}_i$. Further we will discuss only $p$-polarized laser radiation since $s$-polarized light doesn't excite SP. Let us consider the shallow sinusoidal grating $x = h(z)$ with the amplitude $h_0$ and wavenumber $k_g$ (see Fig. 2 (c)):

$$h(z) = h_0 \sin k_g z, \qquad (1)$$
$$\theta(z) \cong k_g h_0 \cos k_g z. \qquad (2)$$

The exact boundary conditions for the tangential and normal projections of the electric field $E_\tau$ and $E_n$ should be written on the wavy surface $x = h(z)$. In the case of shallow grating they can be simplified by transferring to the plane $x = 0$. Perturbation theory with the small parameters $k_g h_0$ gives:

$$\mathbf{E}|_{x=h_0 \sin k_g z} \cong \mathbf{E}|_{x=0} + \frac{\partial \mathbf{E}}{\partial x}\bigg|_{x=0} \cdot h_0 \sin k_g z. \tag{3}$$

Also we should take into account that the tangential and normal projections of the electric field include both $E_z$ and $E_x$ components:

$$E_\tau \cong E_z + E_x \cdot \theta(z), \tag{4.1}$$
$$E_n \cong E_x - E_z \cdot \theta(z). \tag{4.2}$$

Now boundary conditions $E_{\tau 1} = E_{\tau 2}$ and $E_{n1} = \varepsilon E_{n2}$ can be transferred from the surface $x = h(z)$ to the plane $x = 0$ using linear approximations (3), (4.1) and (4.2):

$$E_{z1} - E_{z2} = \left(\frac{\partial E_{z2}}{\partial x} - \frac{\partial E_{z1}}{\partial x}\right) h_0 \sin k_g z + (E_{x2} - E_{x1}) \cdot k_g h_0 \cos k_g z, \tag{5}$$

$$E_{x1} - \varepsilon E_{x2} = (E_{z1} - \varepsilon E_{z2}) \cdot k_g h_0 \cos k_g z + \left(\varepsilon \frac{\partial E_{x2}}{\partial x} - \frac{\partial E_{x1}}{\partial x}\right) h_0 \sin k_g z, \tag{6}$$

where $\varepsilon$ is the permittivity of metal, indexes "1" and "2" refer to vacuum and metal half-space correspondingly. The right sides of Eqns. (5)–(6) describe their differences from "common" boundary conditions on a flat surface; the differences vanish when $h_0 = 0$.

Due to spatial harmonics $e^{\pm i k_g z}$ appeared in the boundary conditions (5)–(6), the reflected optical radiation includes two waves of the first diffraction order. If the incident wave has the form

$$\mathbf{E}_i(x, z, t) = \mathbf{E}_i e^{i\omega t - i k_z z + i k_x x} + c.c., \tag{7}$$

then the electric field of diffraction waves above the metal can be written as

$$\mathbf{E}_{d\pm}(x, z, t) = \mathbf{E}_{d\pm} e^{i\omega t - i k_{z\pm} z - i k_{x\pm} x} + c.c., \tag{8}$$

where $k_{z\pm} = k_z \pm k_g$, $k_{x\pm} = \sqrt{k_0^2 - k_{z\pm}^2}$ and $k_0$ is the wavenumber of the incident wave. We will pay special attention to the case when $k_{x+}$ is imaginary value, so $\mathbf{E}_{d\pm}$ is non-propagating and exponentially decaying in $x > 0$ direction. Together with the diffraction waves (8) we need to consider exponentially decaying "transmitted" electromagnetic modes in the lower half-space $x < 0$ with the same $z$-components $k_{z\pm}$:

$$\mathbf{E}_{t\pm}(x, z, t) = \mathbf{E}_{t\pm} e^{i\omega t - i k_{z\pm} z + \eta_{t\pm} x} + c.c., \tag{9}$$

where $\eta_{t\pm} = \sqrt{k_{z\pm}^2 - \varepsilon k_0^2}$. The full set of electromagnetic modes satisfying conditions (5)–(6) in a first order of perturbation theory is shown in Fig. 1 (b).

Substituting Eqns. (7)–(9) to the boundary conditions (5)–(6) and decomposing them into spatial harmonics $e^{-i k_z z}$, $e^{-i k_{z\pm} z}$ we obtain a linear system of equations. After straightforward algebra we find expressions for all the wave amplitudes at given frequency $\omega$. Particularly, for the $z$-component of electric field $\mathbf{E}_{t+}$ we find

$$E_{t+,z} = k_{x+} h_0 E_{i,z} \left\{ \frac{k_{z+}(\eta^2 + \varepsilon k_x^2) + (i k_z + \eta)(1-\varepsilon) k_{x+} k_g - k_z k_{x+} \varepsilon(\eta - i k_x)}{k_{z+}(i\eta - \varepsilon k_x)\underline{(\eta_{t+} + \varepsilon k_{x+})}} \right\}, \tag{10}$$

where $\eta = \sqrt{k_z^2 - \varepsilon k_0^2}$. The underlined denominator becomes zero in the case of exact synchronism between the diffraction wave $\mathbf{E}_{d+}$ and the surface plasmon having wavenumber $k_p = k_0 \sqrt{\varepsilon/(\varepsilon + 1)}$, (Re $\varepsilon < -1$), when

$$k_{z+} = k_0 \sqrt{\frac{\varepsilon}{\varepsilon + 1}}. \tag{11}$$

Similarly, excitation of SP by another diffraction wave $\mathbf{E}_{d-}$ takes place when $k_{z-} = -k_p$. Here we should emphasize that SP mode was included into the model with the finite metal permittivity $\varepsilon$. So the linear transformation of the incident optical wave to surface plasmons was considered without introducing them artificially. Additional absorption of the optical pulse due to the SP excitation can be estimated by calculating of the total energy in the surface plasmon polariton mode. It consists of the waves $\mathbf{E}_{d+}$ and $\mathbf{E}_{t+}$ propagating above and into the metal correspondingly.

The amplitude of SP is limited by three main factors: Drude losses, finite laser pulse duration and SP diffraction losses. The first one means that the permittivity is a complex value $\varepsilon = \varepsilon' + i\varepsilon''$, so just a real part of the resonance condition (11) can be fulfilled. For example, if $|\varepsilon'| \gg |\varepsilon''|$, at the resonance point we obtain

$$E_{t+,z} \cong -2 \frac{1 + \cos^2 \alpha}{\sin \alpha} \frac{k_0 h_0}{\varepsilon''} E_{i,z}, \tag{12}$$

where $\alpha$ is the incidence angle of the optical wave counted from the surface.

The resonance condition (11) and Eq. (10) are written at some given resonance frequency $\omega$. Let's consider influence of a frequency detuning $\delta\omega$ on the SP amplitude using the simplest plasma-like model of metal dispersion:

$$\varepsilon = 1 - \frac{\omega_p^2}{\omega^2}. \tag{13}$$

At frequency $\omega + \delta\omega$ we find

$$E_{t+,z}(\omega + \delta\omega) \cong -\frac{1 + \cos^2 \alpha}{\sin \alpha (\cos \alpha - 1)} \frac{\omega^4}{\omega_p^4} k_0 h_0 \frac{\omega}{\delta\omega} E_{i,z}(\omega + \delta\omega). \tag{14}$$

In the discussed experiments [5-9] typical parameters are the following: the laser pulse duration $\tau$ is 50–150 fs, central wavelength is 800 nm, resonant incident angle $\alpha$ is from 30 to 60°, the plasma frequency $\omega_p$ is 5–10 times higher than the optical frequency and the grating depth $h_0$ varies from 10 to 90 nm. In this case frequency detuning $\delta\omega$ limits SP amplitude significantly stronger than Drude absorption in Eq. (12). Fourier transform of Eq. (14) gives an equation for SP amplitude evolution $E_{t+,z}(t)$ under the action of quasimonochromatic incident wave with the central frequency $\omega$:

$$\frac{\partial E_{t+,z}(t)}{\partial t} = \frac{1 + \cos^2 \alpha}{\sin \alpha (1 - \cos \alpha)} k_0 h_0 \frac{\omega^4}{\omega_p^4} \omega E_{i,z}(t). \tag{15}$$

Using (15) we can estimate the surface density of SP energy $w_{SP}$:

$$w_{SP} = \frac{(1 + \cos^2 \alpha)^2 (k_0 h_0)^2}{8(1 - \cos \alpha)^2} \frac{\omega_0^5}{\omega_p^5} (\omega_0 \tau)^2 \frac{|\mathbf{E}_i|^2}{2k_0}. \tag{16}$$

A coefficient of optical-to-SP conversion (full energy of the surface plasmon $W_{SP}$ divided by the incident laser pulse energy $W_{opt}$) is the following:

$$\frac{W_{SP}}{W_{opt}} = \sqrt{\frac{\pi}{2}} \frac{(1 + \cos^2 \alpha)^2}{(1 - \cos \alpha)} \frac{\omega_0^5}{\omega_p^5} (k_0 h_0)^2 \omega_0 \tau. \tag{17}$$

Typically in the experiments [5-9] femtosecond laser pulse length is much shorter than the diameter of the beam (15-30 microns vs. 1-10 millimeters correspondingly). The small area of the surface currently irradiated by the laser pulse is moving along the interface with the superluminal phase velocity $v = c/\cos \alpha$ (see Fig. 2). Heating of electrons in the optical skin-layer and further heat diffusion lead to formation of the electronic temperature distribution $T_e(x, z - vt)$ moving along the $z$ axis. We will also use the average kinetic energy of electron $\varepsilon_e$, which is proportional to $T_e^2$ in the case of degenerate electron gas with the Fermi energy $\varepsilon_F \gg T_e$ [15]:

$$\varepsilon_e(x, z - vt) \cong \frac{3}{5}\varepsilon_F + \frac{\pi^2}{4}\frac{T_e^2}{\varepsilon_F}. \tag{18}$$

In the paper [10] it was demonstrated analytically that moving thermal energy distribution $\varepsilon_e(x, z - vt)$ produces low-frequency radiation with the following tangential electric field $E_z$ above the surface of metal:

$$E_z \cong -\frac{2\cos\alpha}{3ce}\frac{\partial \varepsilon_e}{\partial t}\bigg|_{x=0}, \tag{19}$$

where e is the elementary charge.

From Eqns. (18) and (19) it follows that the THz pulse waveform is determined only by the temporal dynamics of the electronic temperature $T_e(t)$ in the vicinity of the surface. Note that the solution (19) is valid for normal metals with high free electron density $n$ (Au, Ag, Cu et al.) for which the plasma frequency $\omega_p = \sqrt{4\pi n e^2/m}$ is much higher than both the electron scattering rate ν and the frequency of THz field $\omega_{THz}$.

Initially Eq. (19) was obtained for the case of flat surface. At the same time, the discussed metal gratings have a typical depth of about 10–100 nm and a period comparable to the laser wavelength, which is 3 orders of magnitude less than the THz wavelength. Therefore, in the analytical calculations we will assume that the presence of grating has no direct influence on the electromagnetic fields at THz frequencies and changes only the structure of optical fields. Namely, in the resonant geometry excitation and absorption of SP at optical frequency can significantly increase the thermal energy $\delta\varepsilon_e$ absorbed by one electron during the laser pulse reflection. According to Eq. (19), this should increase generated THz signal. Note that our numerical results (see below, after the analytical estimations) don't use such assumptions since they are based on the full system of hydrodynamic equations for electron gas and Maxwell equations.

Eq. (17) allows to calculate directly an average thermal energy received by one electron near the surface due to SP absorption in metal. A low-frequency electric field generated by moving $\varepsilon_e$ distribution can be estimated using Eq. (19), so the full energy of THz signal can be found for given geometry of the laser pulse.

Absolute measurements of the THz signal energy generated from the gold foil were made in [7, 8] (the irradiated area was 0.25 cm$^2$, the grating depth was 45 nm, the laser pulse had central wavelength of 785 nm and duration of 150 fs). The authors found that the THz pulse with the energy of 0.42 fJ was generated when the optical fluence was 0.76 mJ/cm$^2$. Since the spectral sensitivity of the THz detector was limited by the range of 0.34–0.38 THz, the full comparison with the presented model is quite complicated. However, for the listed parameters resonant SP excitation provides an additional absorption of about 7% of the incident pulse energy according to Eq. (17), so the SP mode enhance total optical absorption significantly. These parameters correspond to the thermal energy of 6-7 meV per one free electron in the skin-layer. Estimation based on Eq. (19) gives the full THz signal energy of about 1 fJ which fits the available data in order of magnitude. This demonstrates that thermal nonlinearity in the electron gas enhanced by the SP excitation has enough magnitude for explaining the resonant THz response.

We performed numerical study of the discussed mechanism of THz field generation on the basis of hydrodynamic equations for free electrons in metal and Maxwell equations. The computational scheme was similar to the one used in previous studies of optical-to-THz conversion on flat metal surfaces and in arrays of nanoparticles [10, 16]. First of all, numerical modeling demonstrates fairly good agreement with the analytical formula for the generated THz field given by Eq. (19). Modeling results show a resonant character of THz response when the laser pulse incidence angle is varied (similarly, when the grating period $l_g$ is varied – see Fig. 3 c), d)). Terahertz generation efficiency has optimal grating depth $d_g$ as well, this effect will be discussed below. For the case of considerably deep gratings optimal grating period $l_g$ significantly depends on grating shape: compare the cases of sinusoidal and rectangular gratings in Fig. 3 a), b). Optimal period for the sinusoidal grating shows more stability with respect to grating depth and tends to theoretically calculated value (vertical dash-dotted line in Fig. 3a).

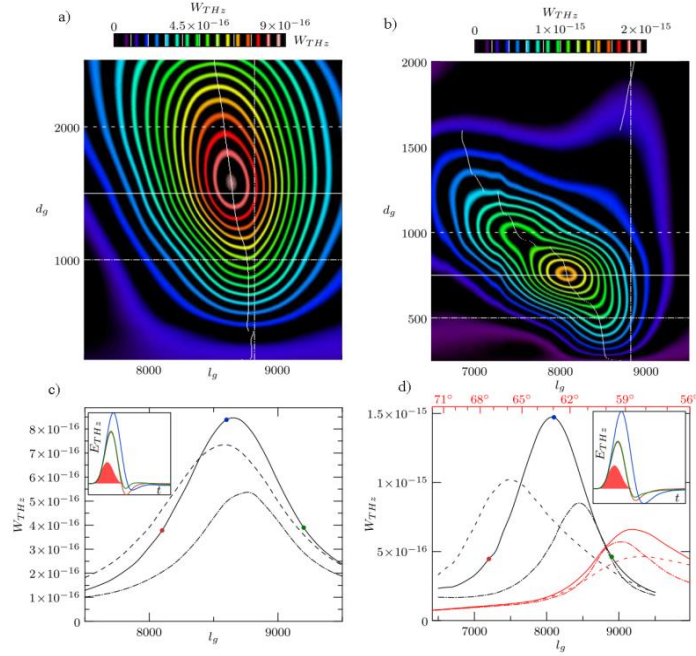

Fig. 3. Dependence of the THz signal energy $W_{THz}$ on the grating period $l_g$ and depth $d_g$ obtained in the numerical modeling a), b). From here on all the values are given in arb. u., the optical wavelength is 4500 arb. u. and the skin-layer thickness is 100 arb. u., incidence angle counted from surface is 60 degrees. Plots c) and d) show corresponding cross-sections marked in a) and b) with corresponding line style; insets show waveforms of the THz pulses corresponding to the colored points; the filled red graph is the laser pulse intensity. Plots a) and c) are for grating of sinusoidal shape; solid, dashed and dot-dashed lines correspond to the grating depths of 1500, 2000 and 1000 arb. u. Plots b) and d) are for grating of rectangular shape; solid, dashed and dot-dashed lines correspond to the grating depths of 750, 1000 and 500 arb. u., red curves depicts angular dependencies of the THz pulse energy at $l_g = 8821$ arb. u.

Another experimental feature which can be interpreted by the proposed generation mechanism is the THz response decrease when the grating depth is too large ($d = 2h_0$ is more than ~100 nm in [8]). For this we should take into account not only the process of incidence wave transformation to SP, but also SP diffraction on the grating (in particular, transformation of SP to the reflected optical wave). When the diffraction losses of SP exceeds Drude losses, the average energy absorbed by one electron decreases despite the further growth of the grating depth. The optimal value of $d$ obtained in the numerical modeling for the rectangular grating (Fig. 4, right panel) is about 14% of the optical wavelength which coincides with the experimental data from the paper [8]. In the Fig. 5 the full parameter space "grating depth/grating period" is presented.

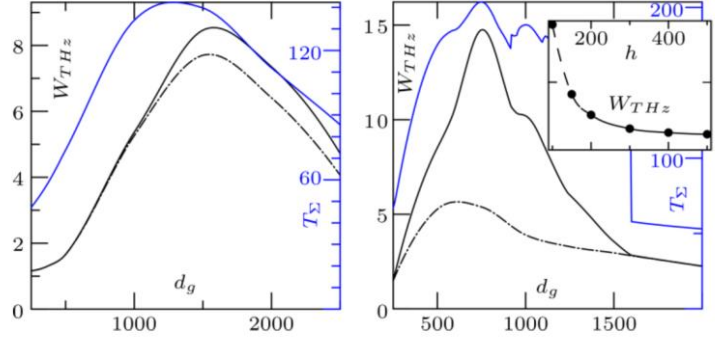

Fig. 4. Black lines – dependence of the THz pulse energy on the grating depth d = 2h0 for sinusoidal a) and rectangular b) gratings. Due to the change of the resonant grating period lg with the depth increase, two types of curves are shown: all the solid lines are plotted for the optimal period at each value of dg and dashed lines are plotted at the fixed period of 8821 arb. u., which is obtained from theoretical condition of efficient SP generation for given parameters. Solid and dot-dashed lines correspond to vertical cross-sections in Fig. 3 a) b). Blue lines show the total thermal energy in arb. u. absorbed by the electronic subsystem. Inset: THz pulse energy vs. thickness of the metal foil hg in arb.u (skin-layer thickness is 100 arb.u.).

In the paper [6] optical-to-THz conversion efficiencies were compared for the gold foil thicknesses of 20, 30, 40 and 50 nm (deposition on the silicon substrate). It was found that the optimal thickness is about 40 nm while using 20 nm or 50 nm foil reduced efficiency by 2-3 times. This dependence cannot be fully explained by the presented model since at 20-30 nm thicknesses the anomalous (ballistic) heat diffusion should be taken into account and also the experimental samples uncontrollably lose continuity. Fig. 5 shows that our model predicts only several times increase of THz signal for thin foils.

To sum up, we showed that laser-induced thermal effects inside the metal lead to the formation of delayed electromagnetic pulse of the THz range. In the case of structured surface, taking into account excitation, absorption and diffraction of SP at optical frequency allow to interpret resonance character of THz response and explain its dependence on the grating depth.

**Acknowledgements**

This work was supported by Russian Foundation for Basic Research (RFBR) grants #17-02-00387 and #18-42-520023. I. O. is grateful to Foundation for the Advancement of Theoretical Physics and Mathematics "BASIS".

**References**

1. Seungchul Kim, Jonghan Jin, Young-Jin Kim, In-Yong Park, Yunseok Kim & Seung-Woo Kim, High-harmonic generation by resonant plasmon field enhancement, Nature volume 453, p. 757–760 (2008)

2. Guillaume Baffou Romain Quidant, Thermo-plasmonics: using metallic nanostructures as nano-sources of heat, Laser & Photonics Reviews Volume 7, Issue 2, (2012)

3. J. Kupersztych, P. Monchicourt, and M. Raynaud, Ponderomotive Acceleration of Photoelectrons in Surface-Plasmon-Assisted Multiphoton Photoelectric Emission, Phys. Rev. Lett. 86, 5180 – (2001)

4. Hou, W. and Cronin, S.B. A Review of Surface Plasmon Resonance Enhanced Photocatalysis. Advanced Functional Materials, 23, 1612-1619, (2013).

5. G. H. Welsh, N. T. Hunt, K. Wynne, Terahertz-pulse emission through laser excitation of surface plasmons in a metal grating, Physical review letters, Vol. 98, №2, P. 026803 (2007).

6. G. H. Welsh, K. Wynne, Generation of ultrafast terahertz radiation pulses on metallic nanostructured surfaces, Optics Express, Vol. 17, № 4, P. 2470 (2009).

7. F. Garwe, A. Schmidt, G. Zieger, T. May, K. Wynne, U. Hübner, M. Zeisberger, W. Paa, H. Stafast, H.-G. Meyer, Bi-directional terahertz emission from gold-coated nanogratings by excitation via femtosecond laser pulses, Applied Physics B. — 2011. — Vol.102, №3. — P. 551—554.

8. A. Schmidt, F. Garwe, U. Hubner et al. Experimental characterization of bi-directional terahertz emission from gold-coated nanogratings, Applied Physics B. — 2012. — Vol.109, №4. — P. 631—642.

9. G. K. P. Ramanandan, G. Ramakrishnan, N. Kumar et al. Emission of terahertz pulses from nanostructured metal surfaces, Journal of Physics D: Applied Physics. — 2014. — Vol.47, №37. — P. 374003.

10. I. V. Oladyshkin, D. A. Fadeev, V. A. Mironov, Thermal mechanism of laser induced THz generation from a metal surface, Journal of Optics. — 2015. — Vol.17, №7. — P. 075502.

11. F. Kadlec, P. Kuzel, J.-L. Coutaz, Optical rectification at metal surfaces, Optics Letters. — 2004. — Vol.29, № 22. — P. 2674—2676

12. F. Kadlec, P. Kuzel, J.-L. Coutaz, Study of terahertz radiation generated by optical rectification on thin gold films, Optics Letters. — 2005. — Vol.30, № 11. — P. 1402—1404.

13. E. V. Suvorov, R. A. Akhmedzhanov, D. A. Fadeev, I. E. Ilyakov, V. A. Mironov, B. V. Shishkin, Terahertz emission from a metallic surface induced by a femtosecond optic pulse, Optics Letters. — 2012. — Vol.37, № 13. — P. 2520.

14. S. G. Bezhanov, S. A. Uryupin, Free-electron mechanisms of low-frequency radiation generation on metal surfaces, Optics Letters. — 2016. — Vol.41, №21. — P. 4975—4978.

15. Pitaevskii L. P. and Lifshitz E. M., Physical Kinetics (Amsterdam: Elsevier), 2012.

16. D. A. Fadeev, I. V. Oladyshkin, V. A. Mironov, Terahertz emission from metal nanoparticle array Optics Letters 43 (8), 1939-1942 (2018)